\documentclass[12pt,a4paper,dvips]{article}
\usepackage{epsfig}
\usepackage{ a4p}
\usepackage{cite,mcite}
\usepackage{graphicx}
\usepackage{ physics, l3_title,ifthen, Lep}
\journalname{Phys. Lett. B}
\date{December 22, 2000}
\preprint{2000-155}
\Lep{2}
\newlength{\capindent}
\newlength{\capwidth}
\newlength{\figwidth}
\newcommand{\icaption}[2][!*!,!]{\hspace*{\capindent}%
  \begin{minipage}{\capwidth}
    \ifthenelse{\equal{#1}{!*!,!}}%
      {\caption{#2}}%
      {\caption[#1]{#2}}
  \end{minipage}}

\def\NPB{{Nucl. Phys.} {\bf B }}
\def\PLB{{Phys. Lett.} {\bf B }}

\def\PRD{{Phys. Rev.} {\bf D }}
\def\ZPC{{Z. Phys.} {\bf C }}
\def\CPC{Comp. Phys. Comm. }

\def\ra{\rightarrow}

\def\eeeeh{$\mathrm{e^+e^-} \rightarrow \mathrm{e^+e^-}${\sl hadrons}}
\def\ccarx{\ensuremath{\mathrm{c \bar{c}X}}}

\def\Wgge {W_{\gamma \gamma}}
\def\Wgg {$W_{\gamma \gamma }$}
\def\Wvis {$W_{\mathrm{vis}}$}

\def\Wvise {W_{\mathrm{vis}}}
\def\Lgg  {$\cal{L}_{\gamma \gamma }$}

\def\seehe { \Delta \sigma ( \ee \ra   \ee \ccarx)}
\begin{document}
\bibliographystyle{l3style}
\begin{titlepage}
\title{Measurement of the Charm Production Cross Section   \\
       in \boldmath{${\gamma \gamma}$} Collisions at LEP}

\author{The L3 Collaboration}

%
%

\begin{abstract}
  Open charm production in $\gamma \gamma$ collisions is studied with
data collected at $\rm{e^+ e^-}$ centre-of-mass energies from
189 GeV to 202 GeV corresponding to a total
integrated luminosity of 410 pb$^{-1}$.
The charm cross section   
$\sigma (\gamma\gamma \rightarrow \rm{c \bar{c} X}$) is measured 
for the first time
as a function of the two-photon centre-of-mass energy in the 
interval from 5~GeV to 70~GeV
and is compared to NLO QCD calculations. 

\end{abstract}
%
%
\submitted

\end{titlepage}

\begin{figure}[htbp]
\begin{center}
 \mbox{\epsfig{file=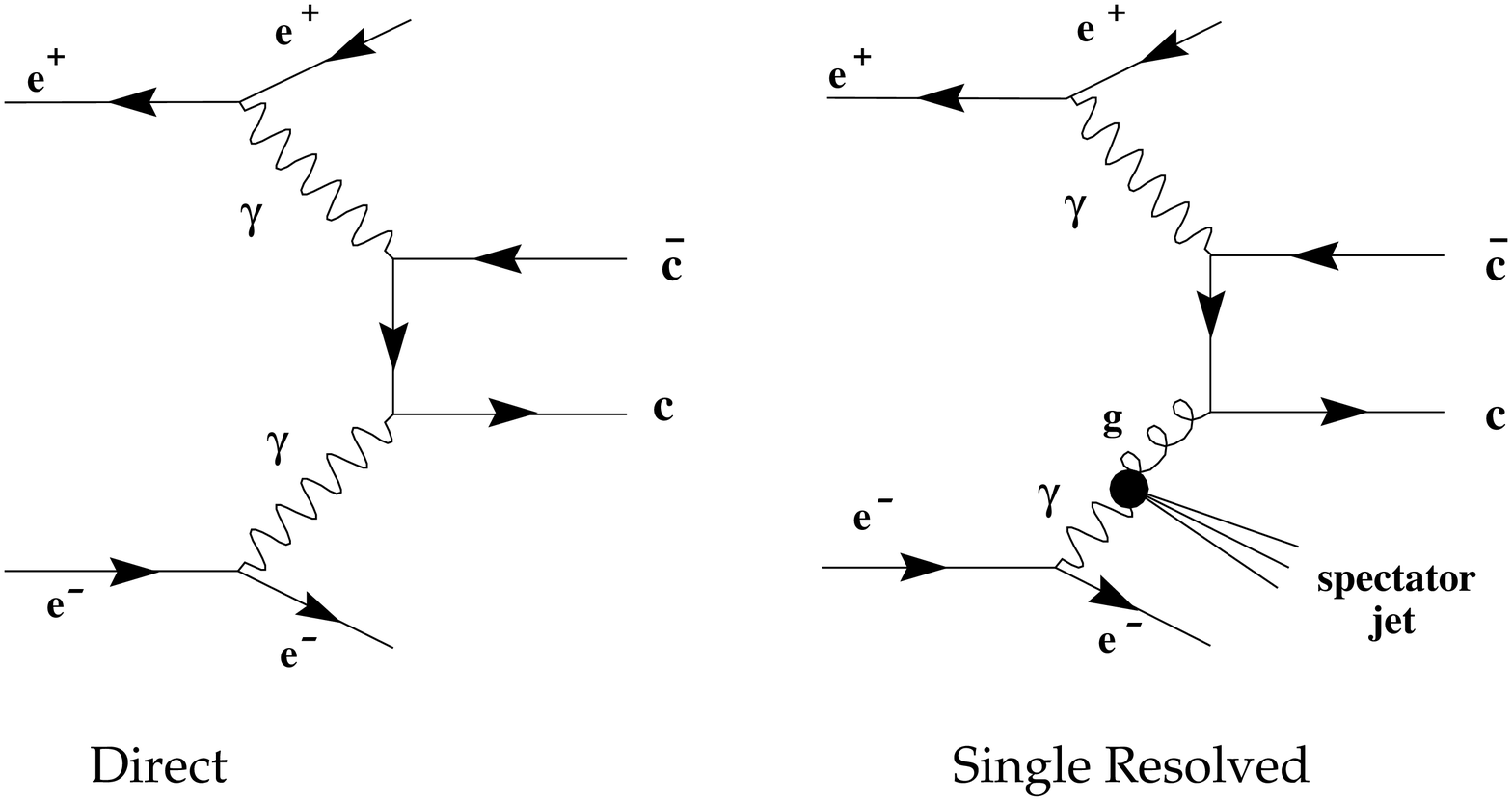, width=.9\textwidth}}  
  \caption{Diagrams contributing to charm
production in $\mathrm{\gamma \gamma}$
    collisions.}
  \label{fig:Feynman}
\end{center}
\end{figure}

\section{Introduction}

\ 
 The cross section for the interaction of two quasi-real photons 
is described by three components. The first ``soft'' component 
is described by  
the Vector Dominance Model (VDM), 
parametrized in hadronic phenomenology by Regge poles.
A second ``direct'' component, 
the point like reaction $\gamma \gamma \rightarrow \rm{q \bar{q}}$, 
can be calculated in QED. Finally, there is a ``hard component"
(resolved or anomalous QCD component)
which requires knowledge of the quark and gluon parton density
functions of the photon. 
 The direct and hard contributions to two-photon 
interactions can be measured in open heavy flavour production.
 In this case a hard physical scale is given by
the c- or b-quark mass. At LEP
energies, the direct and single resolved processes,
shown in Figure~\ref{fig:Feynman}, are predicted to give comparable
contributions to the charm production cross section \cite{theory}, whereas 
at low energies the direct process dominates.
The main contribution to the resolved photon
cross section is the photon-gluon fusion process 
$\mathrm{\gamma g \rightarrow  c \bar{c}}$.
Contributions to charm production arising 
from VDM processes and from doubly 
resolved processes are expected to be small\cite{theory}.

This letter presents the measurement of the
$\mathrm{\gamma \gamma \rightarrow c \bar{c} X}$
 cross section as a function of the two-photon centre-of-mass energy 
$W_{\gamma\gamma}$ in the 
interval 5~GeV~$\leq W_{\gamma\gamma} \leq$~70~GeV.
The data correspond to a total integrated
luminosity $\mathrm{\cal{L} =}$ 410 pb$^{-1}$, 
collected with the L3 detector~\cite{L3det} at centre-of-mass energies 
$\sqrt{\rm{s}} = 189 - 202$~GeV.
The inclusive charm production
cross section $\sigma(\rm{e}^{+} \rm{e}^{-} \rightarrow 
 \rm{e}^{+} \rm{e}^{-} c \bar{c} X$) was measured 
by L3 at $\sqrt{\rm{s}} = 91 - 202$~GeV~\cite{L3cc,L3ccbb}.
Charm quarks are identified by their decay 
into electrons\footnote{Electron stands for electron or positron throughout this
paper.} ~\cite{L3ccbb}.

\section{Monte Carlo}

The PYTHIA \cite{pythia} Monte Carlo is
used to model the \eeeeh {} processes.
Quarks other than b are taken as massless 
in the corresponding matrix elements~\cite{massless}. 
The resolved process uses the SaS1d photon
structure function \cite{sas1d}.   
The two-photon luminosity function is implemented in the 
equivalent photon approximation (EPA)~\cite{Budnev} with a 
cutoff $\mathrm{Q^2 < m_{\rho}^2}$, where $m_{\rho}$ 
is the mass of the $\rho$ meson. 

The background sources are 
$\mathrm{e^{+}e^{-} \ra e^{+}e^{-} \tau^{+} \tau^{-}}$, 
$\mathrm{e^{+}e^{-} \ra q \bar{q}}$, 
$\mathrm{e^{+}e^{-} \ra \tau^{+} \tau^{-}}$
and $\mathrm{e^{+}e^{-}} \ra \mathrm{W^{+} W^{-}}$.
These processes are generated by JAMVG~\cite{verm}, PYTHIA, 
KORALZ~\cite{koralz} and KORALW~\cite{koralw} respectively. 
The detector simulation is performed using the 
GEANT \cite{GEANT} and GHEISHA \cite{GHEISHA} packages. 
The Monte Carlo
events are reconstructed in the same way as the data. 
Time dependent detector inefficiencies, 
as monitored during the data taking period,
are also simulated.

\section{Measurement of \boldmath{$\sigma(\mathrm{\gamma \gamma \rightarrow c \bar{c} X})$}}

The measurement of the $\sigma(\mathrm{\gamma \gamma \rightarrow c \bar{c} X})$ cross 
section 
is performed on the high statistics electron sample
of 2434 events discussed in
Reference \cite{L3ccbb}. 
The backgrounds from 
annihilation processes and two-photon production of tau pairs 
are estimated to be 0.75\% and are subtracted from the data. 
The background from $\mathrm{\gamma \gamma \rightarrow b \bar{b} X}$ 
events is modelled with
the PYTHIA Monte Carlo assuming equal contributions
from direct and resolved processes. It is also subtracted
from the data assuming our measured cross section~\cite{L3ccbb}.
The charm purity after background subtraction is 75\%.

\begin{figure}[tbp]
\begin{center}
 \mbox{\epsfig{file=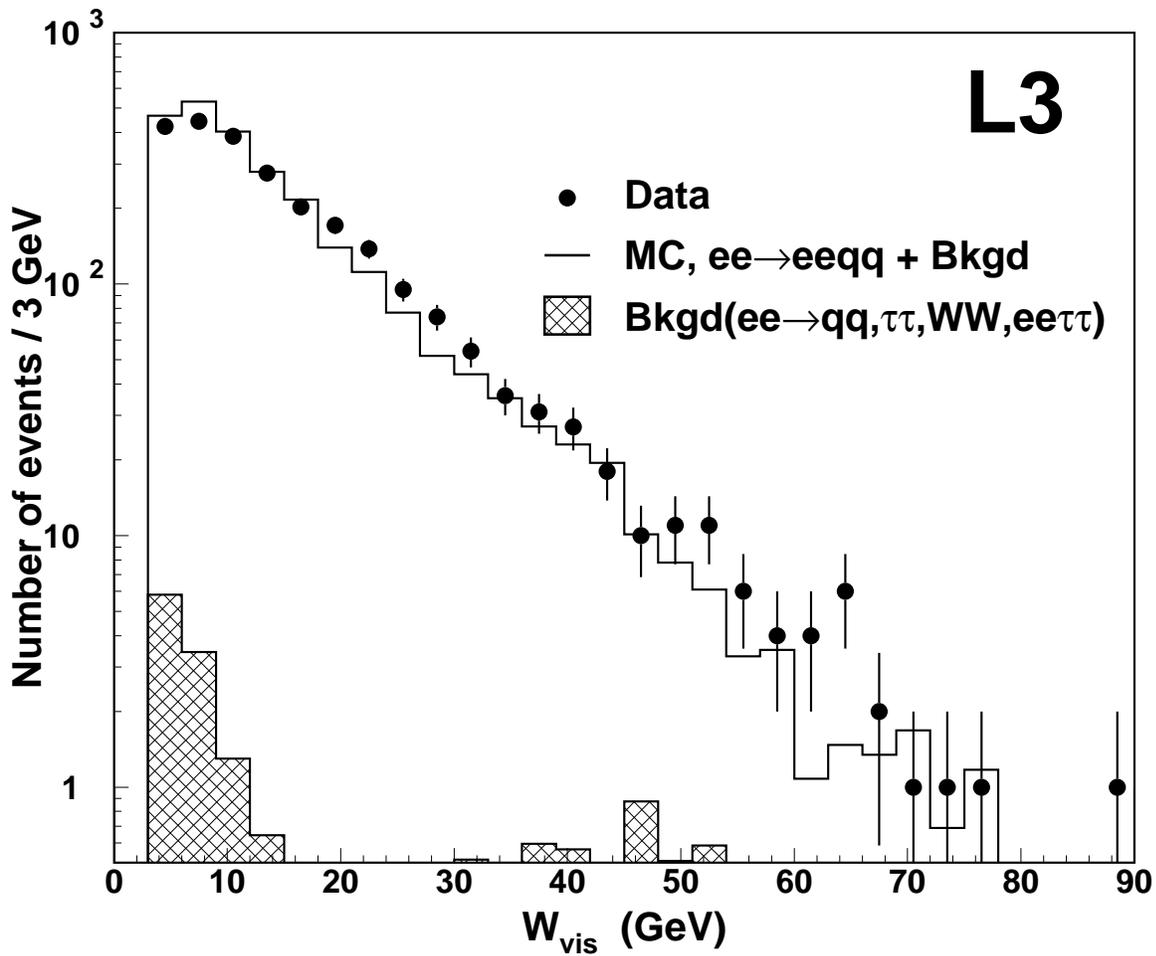, width=.95\textwidth}}  
  \caption{The visible mass spectrum, $W_\mathrm{vis}$, 
for the inclusive electron data at 
  $\sqrt{s}=189-202$ GeV compared to the PYTHIA prediction.
The Monte Carlo spectrum with all 
flavour contributions is normalized to the integrated
luminosity of the data after scaling to the measured charm and 
beauty cross sections according to our measurements~\protect\cite{L3ccbb}. 
Other background sources are also shown.}
  \label{fig:wvis9899}
\end{center}
\end{figure}

The analysis is restricted to events 
with visible mass $W_\mathrm{vis}>3$ GeV.
$W_\mathrm{vis}$ 
is calculated from the four-momentum vectors of the
measured particles, tracks and calorimetric clusters
including those from the small angle luminosity monitor.
The $W_\mathrm{vis}$ distribution is corrected for trigger efficiency 
using data with a set of independent triggers. 
It varies from 94$\%$ at $W_\mathrm{vis} =3$ GeV to
98$\%$ at $W_\mathrm{vis} \geq 30$ GeV.
Table~\ref{tab:numbevents} shows the selected data events in bins of \Wvis {}
and the different contributions of the signal
and the background predicted by the PYTHIA Monte Carlo. 
A comparison of the
visible mass distribution of the data after final selection
with the expectations of the PYTHIA Monte
Carlo at ${\sqrt{s}=189-202}$ GeV is shown in
Figure~\ref{fig:wvis9899}. 
  
\subsection{Unfolding procedure }

\begin{figure}[tbp]
\begin{center}
 \mbox{\epsfig{file=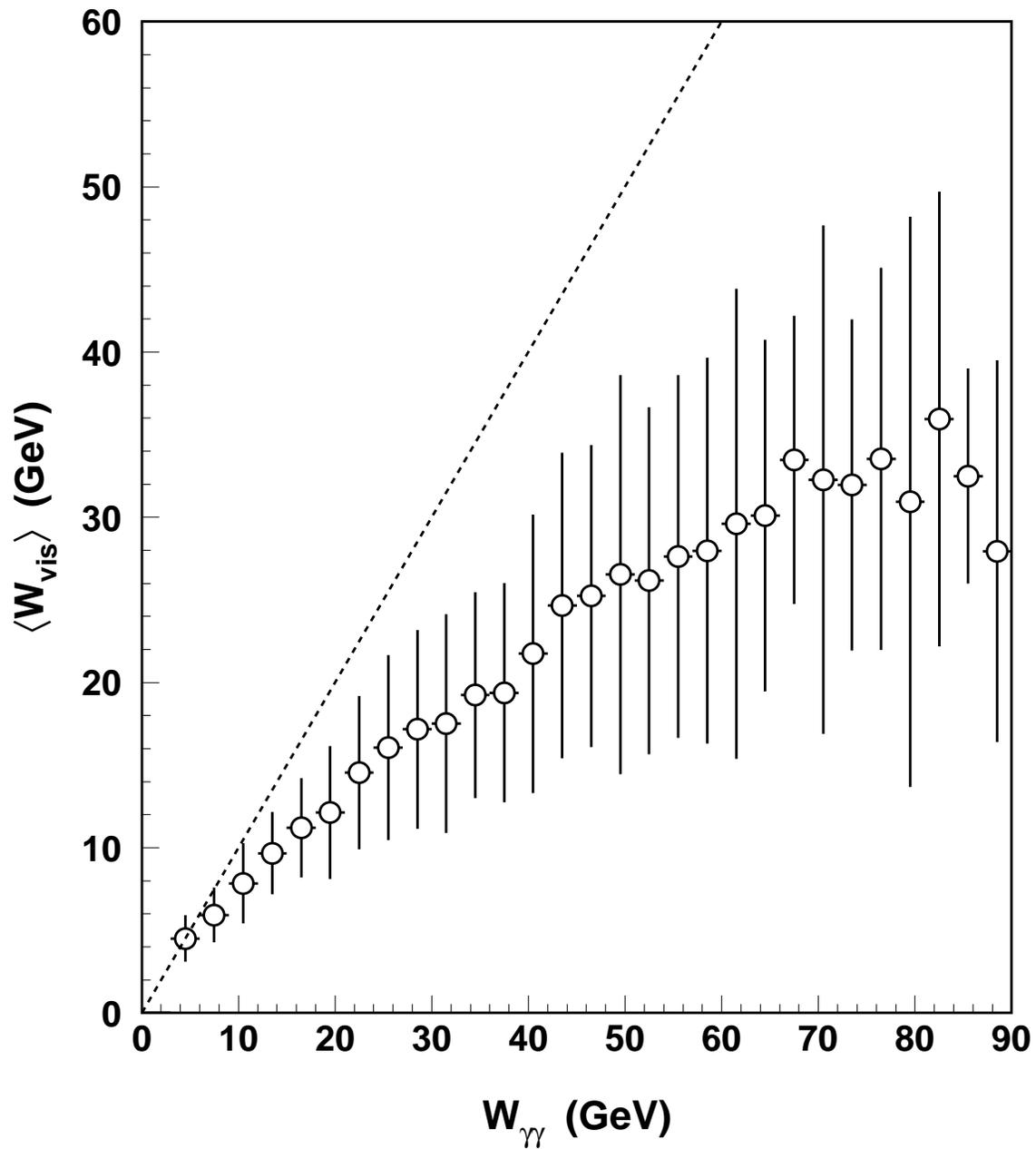, width=.95\textwidth}}  
  \caption{Profile of the distribution of $W_\mathrm{vis}$ as a function
of $W_\mathrm{\gamma\gamma}$ using the PYTHIA
Monte Carlo. The open circles correspond to the mean values and
the error bars represent the r.m.s. of the distribution.
The dashed line corresponds to perfect correlation.}
  \label{fig:wvisvswgg}
\end{center}
\end{figure}

The hadronic final state is not always fully contained 
in the detector acceptance. 
An unfolding procedure is hence applied to obtain 
the $\gamma \gamma$ centre-of-mass 
energy $W_\mathrm{\gamma \gamma}$ from $W_\mathrm{vis}$.
Figure~\ref{fig:wvisvswgg} shows the correlation of 
the $W_\mathrm{vis}$ average value, $\langle W_\mathrm{vis} \rangle$, with 
$W_\mathrm{\gamma\gamma}$ as predicted by the Monte Carlo. 
The unfolding corrects for both missing particles
and detector resolution by the relation:
\begin{equation}
 N(\Wgge (i))  = 
 \sum_{j} A_{ij} N(\Wvise (j)),
\end{equation}
where $N(\Wgge (i))$ is the content of the $i$-th interval of $W_\mathrm{\gamma \gamma}$
listed in Table~\ref{tab:cross}, and $ N(\Wvise (j))$ is 
the content of the $j$-th
interval of $W_\mathrm{vis}$ listed in Table~\ref{tab:numbevents}.
 The matrix $A_{ij}$ is 
constructed 
by  considering for each Monte Carlo event  
its reconstructed \Wvis {} and generated \Wgg ~\cite{dagostini}:
\begin{equation}
  A_{ij} = \frac{ P(\Wvise (j)|\Wgge (i)) 
  P(\Wgge (i))}
  { \sum_{l} P(\Wvise (j)|\Wgge (l))
   P(\Wgge (l))},
\end{equation}
where $ P(\Wvise  | \Wgge)$ 
is the likelihood of observing the measured value \Wvis {} given a
generated value of \Wgg {} and  $P(\Wgge)$ is the generated \Wgg {} distribution
 after the selection cuts.
\par
The unfolding matrix is determined using charm events generated with PYTHIA. 
The unfolding uncertainty is estimated with PYTHIA 
using five flavour events instead of charm events and using 
the PHOJET\cite{phojet} Monte Carlo generator with all quark flavours.
This comparison leads to an estimated uncertainty of 5\%.
Charm production in the PHOJET generator is implemented
with a high $p_t$ threshold which prevents the use of PHOJET for
charm efficiency studies.

After unfolding, the events are corrected for  
efficiency using the ratio between selected and 
generated charm events in each  \Wgg {} interval. 
The events with 3~GeV~$\leq W \mathrm{_{vis}} \leq$~5~GeV 
and $W\mathrm{_{vis}} \geq$~70~GeV
are used for unfolding,
 but are excluded from the measurement due to 
their large correction factors and unfolding uncertainty.

\subsection{Cross section measurement}

The cross section 
$\Delta \sigma (\rm{e^+ e^- \rightarrow e^+ e^- c \bar{c} X})$ is measured 
from the number of  events corrected for the efficiency
with the PYTHIA Monte Carlo in each $W_{\gamma \gamma}$ interval 
and the integrated luminosity.
 \par To  extract the cross section $\sigma (\gamma\gamma \rightarrow \rm{c \bar{c} X})$ 
of two real photons, the photon flux \Lgg~\cite{Budnev} is calculated and 
the hadronic two-photon process is extrapolated to $Q^2  = 0$, 
where $Q^2$ is the virtuality of a photon.
This is done through the following relation:
 \begin{eqnarray*}
  \seehe  =
 \int  \frac{dQ_{1}^{2}} {Q_{1}^{2}} \frac{dQ_{2}^{2}} {Q_{2}^{2}}
\mathcal{L}_{\gamma \gamma } (W_{\gamma \gamma},Q_{1}^{2},Q_{2}^{2}) 
\sigma_{\gamma\gamma \rightarrow \rm{c \bar{c} X}}(W_{\gamma \gamma})
d W_{\gamma \gamma}.
 \end{eqnarray*}
For each \Wgg {} interval  a numerical integration is performed 
over its width and the unmeasured $Q^{2}$ 
of the scattered electrons. 

Table~\ref{tab:cross} gives the efficiencies and
the cross sections $\Delta \sigma (\rm{e^+e^-} \rightarrow \rm{e^+e^- c \bar{c} X})$
and $\sigma (\gamma\gamma \rightarrow \rm{c \bar{c} X})$ as a 
function of $W_\mathrm{\gamma \gamma}$.
The systematic
uncertainties arise mainly from the charm purity estimate, followed 
by the unfolding procedure, the charm efficiency, the beauty cross section~\cite{L3ccbb}, 
the photon flux and the trigger efficiency, all summarized in Table~\ref{tab:systematics}.
The photon flux uncertainty is estimated by the comparison 
of the two models~\cite{PDG2k,Budnev}.
The uncertainty due to the other background processes is negligible. 
Due to the unfolding, the data points are correlated. The correlation matrix is given in 
Table~\ref{tab:correlation}.

\section{Comparison with theory and interpretation }

\begin{figure}[tbp]
\begin{center}
 \mbox{\epsfig{file=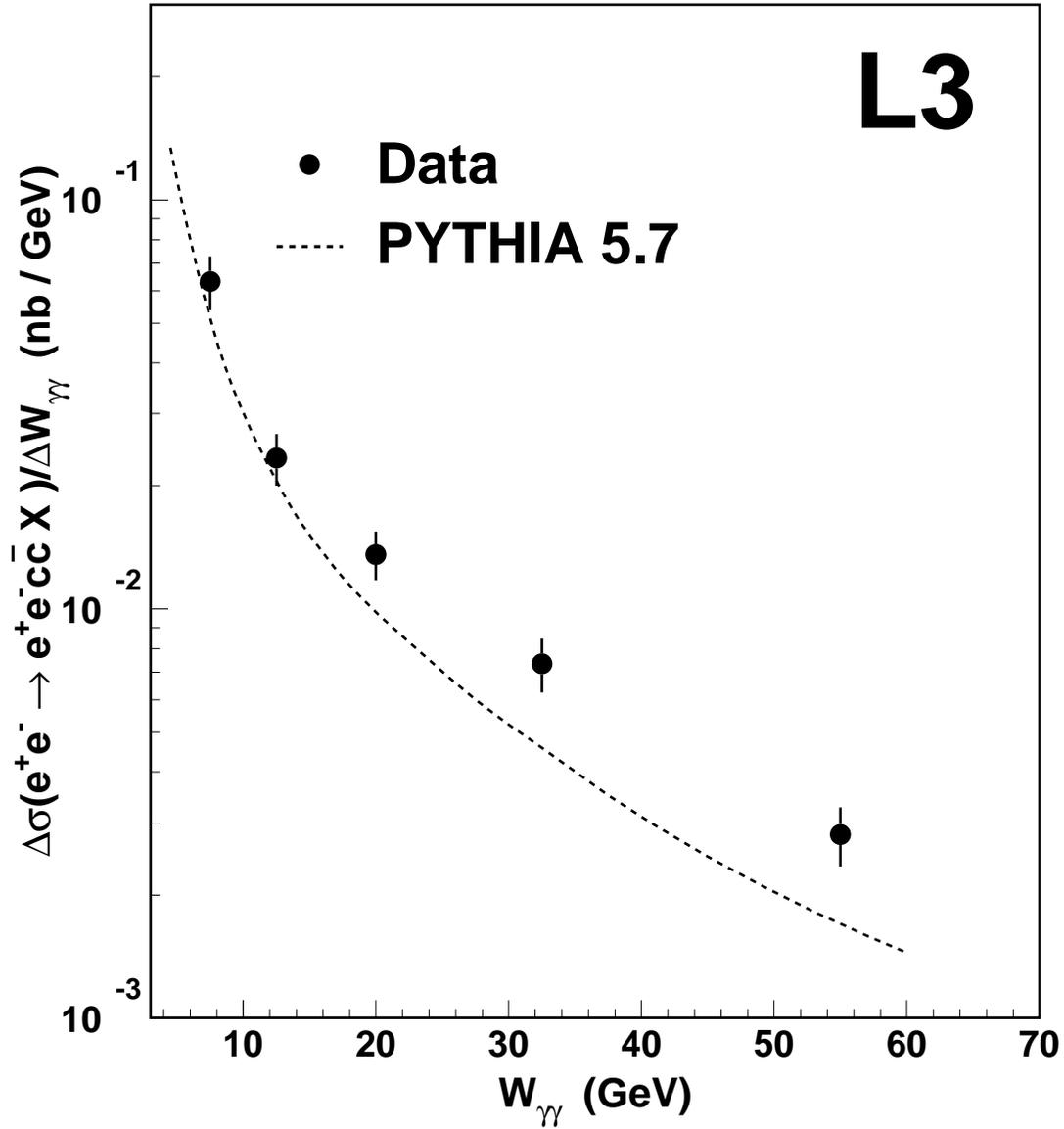, width=.9\textwidth}}  
  \caption{The  
cross section $\mathrm{\Delta \sigma (e^+ e^- \rightarrow e^+ e^- c \bar{c} X}) / \Delta W_\mathrm{\gamma \gamma}$
as a function of $W_\mathrm{\gamma \gamma}$ at $\sqrt{s}=189-202$ GeV.
The dashed line corresponds to the leading order PYTHIA
Monte Carlo prediction.}
  \label{fig:dsigdwgg}
\end{center}
\end{figure}

Figure~\ref{fig:dsigdwgg} shows the differential cross section 
$\Delta \sigma (\mathrm{e^+ e^- \rightarrow e^+ e^- c \bar{c} X}) / \Delta
 W_{\gamma \gamma}$ 
as a function
of $W_{\gamma\gamma}$ measured in 
the interval 5~GeV~$\leq W_{\gamma\gamma} \leq$~70~GeV.
The expected slope from the PYTHIA Monte Carlo 
is steeper than that of the data.

\begin{figure}[tbp]
\begin{center}
 \mbox{\epsfig{file=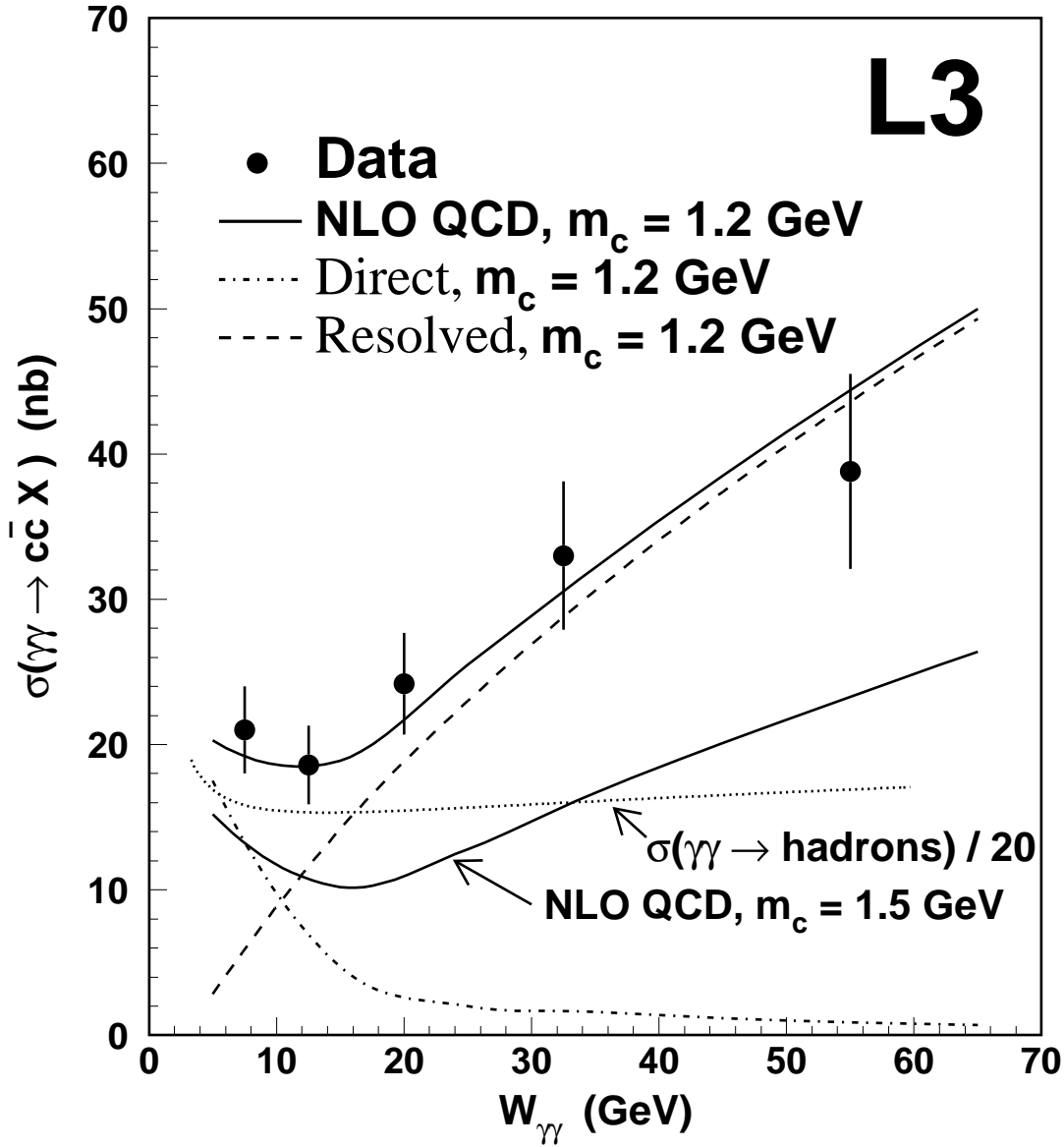, width=.9\textwidth}}  
  \caption{The cross section $\mathrm{\sigma (\gamma \gamma \rightarrow c \bar{c} X)}$
as a function of $W_\mathrm{\gamma \gamma}$ at $\sqrt{s}=189-202$ GeV.
The dotted curve is the 
total cross section $\sigma (\gamma\gamma \rightarrow $ {\sl hadrons}) 
measured by L3~\protect\cite{sigtot} scaled by an arbitrary factor $1/20$.
The continuous line is the NLO QCD prediction, while 
the dashed-dotted and dashed curves show the expectation from 
the direct and resolved process respectively.} 
  \label{fig:sig_ggcc_qcd}
\end{center}
\end{figure}

Figure~\ref{fig:sig_ggcc_qcd} compares the measured cross section 
$\mathrm{\sigma (\gamma\gamma \rightarrow \rm{c \bar{c} X}})$
as a function of $W_{\gamma \gamma}$ with NLO QCD 
calculations~\cite{frixione}. The calculations use
massive quarks in the matrix elements.
The charm mass, $m_\mathrm{c}$, is fixed to 1.2 GeV, 
 the renormalization and
factorization scales are set to $m_\mathrm{c}$ and $2 m_\mathrm{c}$, 
respectively, the QCD parameter $\Lambda^\mathrm{QCD}_5$ is set at 227.5 MeV, 
and the GRS-HO~\cite{GRS} photon 
parton density function is used. 
Using this set of input parameters, the NLO QCD predictions
reproduce well the energy dependence and the normalization.
The calculation with $m_\mathrm{c} = 1.5$ GeV results in about 50\% lower
cross section values, except the first point, where it is lower by 25\%.
A change in the renormalization scale from $m_\mathrm{c}$ to $2 m_\mathrm{c}$
decreases the QCD prediction by 10\% and 30\% 
at low and high $W_{\gamma \gamma}$ respectively.

We compare also the measured charm cross section with the total cross section
of hadron production in two-photon collisions~\cite{sigtot},
scaled by an arbitrary factor $1/20$. 
The slope of $\sigma (\gamma\gamma \rightarrow \mathrm{c \bar{c} X}$) 
is clearly larger than that
of $\sigma (\gamma\gamma \rightarrow $ {\sl hadrons}).

A more quantitative comparison results from fits to the data.
A parametrisation~\cite{DL} 
of the form  
$\sigma = A \, s^{\epsilon} \, + \, B \, s^{-\eta}$, 
``Pomeron $+$ Reggeon'', with $s = W_{\gamma\gamma}^2$ 
describes well the energy behaviour of 
all the total hadron-hadron cross sections 
with universal values $\epsilon = 0.093 \pm 0.002$ and 
$\eta = 0.358 \pm 0.015$\cite{PDG2k}.
A fit to our data with $\epsilon$, $A$ and $B$ being free parameters and 
with fixed $\eta = 0.358$ yields in the interval 5~GeV~$\leq W_{\gamma\gamma} \leq$~70~GeV:
\begin{center}
\begin{tabular}{l}
$\epsilon \phantom{../d.o.f.}= \phantom{0}0.40 \pm 0.03 \phantom{0}(\mathrm{stat.}) \pm \phantom{0}0.07 \phantom{0}(\mathrm{syst.})$ \\
$A \phantom{./d.o.f.}= \phantom{0}1.3 \phantom{0}\pm 0.3 \phantom{00}(\mathrm{stat.}) \pm \phantom{0}0.7 \phantom{00}(\mathrm{syst.}) \phantom{0}\mathrm{nb}$ \\
$B \phantom{./d.o.f.}= 44.0 \phantom{0}\pm 3.8 \phantom{00}(\mathrm{stat.}) \pm 11.1 \phantom{00}(\mathrm{syst.}) \phantom{0}\mathrm{nb}$ \\
$\chi^2 / \mathrm{d.o.f.}\phantom{.} = \phantom{0}8.9 / 2$ \\
\end{tabular}
\end{center}
Correlations between the data points are taken into account.
The systematic uncertainty on 
each parameter is estimated by performing fits with total uncertainties 
and statistical uncertainties only in $\chi^2$.  The systematic uncertainties
are then evaluated by subtracting in quadrature the uncertainty of
the statistical only fit from that given by the total uncertainties.
The fitted value of $\epsilon$ is higher than the universal value~\cite{PDG2k}.

Modifications to the parametrization of the total cross section
were recently proposed~\cite{hardpom}.
In order to allow for a larger slope 
for the high energy behaviour, an admixture of  
``hard'' and ``soft'' Pomerons with 
the Pomeron powers $\epsilon$ of 0.4 and 0.1
respectively was introduced.
Our data correspond to a direct measurement
of the ``hard'' component of photon-photon 
collisions.

A fit to the data by the form  
$\sigma = A \, s^{\epsilon}$,
performed in the interval 10~GeV~$\leq W_{\gamma\gamma} \leq$~70~GeV, yields:
\begin{center}
\begin{tabular}{l}
$\epsilon \phantom{...00000}= 0.26 \pm 0.02 \phantom{0}(\mathrm{stat.}) \pm 0.03 \phantom{0}(\mathrm{syst.})$ \\
$A \phantom{..00000}= 5.0 \phantom{0}\pm 0.4 \phantom{00}(\mathrm{stat.}) \pm 0.8 \phantom{00}(\mathrm{syst.}) \phantom{0}\mathrm{nb}$ \\
$\chi^2 / \mathrm{d.o.f.} = 1.9 / 2$ \\
\end{tabular}
\end{center}

The value of $\epsilon$ is in
agreement with a similar fit to the measurements of the  
photoproduction of J mesons at HERA~\cite{H1, ZEUS}.

\section*{Acknowledgements}
We thank S. Frixione, E. Laenen and
M. Kr\"{a}mer for providing us with NLO QCD calculations
for the process of interest and G.A. Schuler 
for the numerical integration program of the photon flux.
We wish to express our gratitude to the CERN accelerator divisions for the 
excellent performance of the LEP machine. We also acknowledge 
and appreciate 
the effort of the engineers, technicians and support staff who have
participated in the construction and maintenance of this experiment.

\clearpage
\newpage
%
%
\section*{Author List}
\typeout{   }     
\typeout{Using author list for paper 229 -- ? }
\typeout{$Modified: Nov 18 2000 by smele $}
\typeout{!!!!  This should only be used with document option a4p!!!!}
\typeout{   }
%
%
%
%
%
%

\newcount\tutecount  \tutecount=0
\def\tutenum#1{\global\advance\tutecount by 1 \xdef#1{\the\tutecount}}
\def\tute#1{$^{#1}$}
\tutenum\aachen            
\tutenum\nikhef            
\tutenum\mich              
\tutenum\lapp              
\tutenum\basel             
\tutenum\lsu               
\tutenum\beijing           
\tutenum\berlin            
\tutenum\bologna           
\tutenum\tata              
\tutenum\ne                
\tutenum\bucharest         
\tutenum\budapest          
\tutenum\mit               
\tutenum\debrecen          
\tutenum\florence          
\tutenum\cern              
\tutenum\wl                
\tutenum\geneva            
\tutenum\hefei             
\tutenum\seft              
\tutenum\lausanne          
\tutenum\lecce             
\tutenum\lyon              
\tutenum\madrid            
\tutenum\milan             
\tutenum\moscow            
\tutenum\naples            
\tutenum\cyprus            
\tutenum\nymegen           
\tutenum\caltech           
\tutenum\perugia           
\tutenum\cmu               
\tutenum\prince            
\tutenum\rome              
\tutenum\peters            
\tutenum\potenza           
\tutenum\riverside         
\tutenum\salerno           
\tutenum\ucsd              
\tutenum\sofia             
\tutenum\korea             
\tutenum\alabama           
\tutenum\utrecht           
\tutenum\purdue            
\tutenum\psinst            
\tutenum\zeuthen           
\tutenum\eth               
\tutenum\hamburg           
\tutenum\taiwan            
\tutenum\tsinghua          

{
\parskip=0pt
\noindent
{\bf The L3 Collaboration:}
\ifx\selectfont\undefined
 \baselineskip=10.8pt
 \baselineskip\baselinestretch\baselineskip
 \normalbaselineskip\baselineskip
 \ixpt
\else
 \fontsize{9}{10.8pt}\selectfont
\fi
\medskip
\tolerance=10000
\hbadness=5000
\raggedright
\hsize=162truemm\hoffset=0mm
\def\r{\rlap,}
\noindent

M.Acciarri\r\tute\milan\
P.Achard\r\tute\geneva\ 
O.Adriani\r\tute{\florence}\ 
M.Aguilar-Benitez\r\tute\madrid\ 
J.Alcaraz\r\tute\madrid\ 
G.Alemanni\r\tute\lausanne\
J.Allaby\r\tute\cern\
A.Aloisio\r\tute\naples\ 
M.G.Alviggi\r\tute\naples\
G.Ambrosi\r\tute\geneva\
% ...
\newpage
\begin{list}{A}{\itemsep=0pt plus 0pt minus 0pt\parsep=0pt plus 0pt minus 0pt
                \topsep=0pt plus 0pt minus 0pt}
\item[\aachen]
 I. Physikalisches Institut, RWTH, D-52056 Aachen, FRG$^{\S}$\\
 III. Physikalisches Institut, RWTH, D-52056 Aachen, FRG$^{\S}$
\item[\nikhef] National Institute for High Energy Physics, NIKHEF, 
     and University of Amsterdam, NL-1009 DB Amsterdam, The Netherlands
\item[\mich] University of Michigan, Ann Arbor, MI 48109, USA
\item[\lapp] Laboratoire d'Annecy-le-Vieux de Physique des Particules, 
     LAPP,IN2P3-CNRS, BP 110, F-74941 Annecy-le-Vieux CEDEX, France
\item[\basel] Institute of Physics, University of Basel, CH-4056 Basel,
     Switzerland
\item[\lsu] Louisiana State University, Baton Rouge, LA 70803, USA
\item[\beijing] Institute of High Energy Physics, IHEP, 
  100039 Beijing, China$^{\triangle}$ 
\item[\berlin] Humboldt University, D-10099 Berlin, FRG$^{\S}$
\item[\bologna] University of Bologna and INFN-Sezione di Bologna, 
     I-40126 Bologna, Italy
\item[\tata] Tata Institute of Fundamental Research, Bombay 400 005, India
\item[\ne] Northeastern University, Boston, MA 02115, USA
\item[\bucharest] Institute of Atomic Physics and University of Bucharest,
     R-76900 Bucharest, Romania
\item[\budapest] Central Research Institute for Physics of the 
     Hungarian Academy of Sciences, H-1525 Budapest 114, Hungary$^{\ddag}$
\item[\mit] Massachusetts Institute of Technology, Cambridge, MA 02139, USA
\item[\debrecen] KLTE-ATOMKI, H-4010 Debrecen, Hungary$^\P$
\item[\florence] INFN Sezione di Firenze and University of Florence, 
     I-50125 Florence, Italy
\item[\cern] European Laboratory for Particle Physics, CERN, 
     CH-1211 Geneva 23, Switzerland
\item[\wl] World Laboratory, FBLJA  Project, CH-1211 Geneva 23, Switzerland
\item[\geneva] University of Geneva, CH-1211 Geneva 4, Switzerland
\item[\hefei] Chinese University of Science and Technology, USTC,
      Hefei, Anhui 230 029, China$^{\triangle}$
\item[\lausanne] University of Lausanne, CH-1015 Lausanne, Switzerland
\item[\lecce] INFN-Sezione di Lecce and Universit\`a Degli Studi di Lecce,
     I-73100 Lecce, Italy
\item[\lyon] Institut de Physique Nucl\'eaire de Lyon, 
     IN2P3-CNRS,Universit\'e Claude Bernard, 
     F-69622 Villeurbanne, France
\item[\madrid] Centro de Investigaciones Energ{\'e}ticas, 
     Medioambientales y Tecnolog{\'\i}cas, CIEMAT, E-28040 Madrid,
     Spain${\flat}$ 
\item[\milan] INFN-Sezione di Milano, I-20133 Milan, Italy
\item[\moscow] Institute of Theoretical and Experimental Physics, ITEP, 
     Moscow, Russia
\item[\naples] INFN-Sezione di Napoli and University of Naples, 
     I-80125 Naples, Italy
\item[\cyprus] Department of Natural Sciences, University of Cyprus,
     Nicosia, Cyprus
\item[\nymegen] University of Nijmegen and NIKHEF, 
     NL-6525 ED Nijmegen, The Netherlands
\item[\caltech] California Institute of Technology, Pasadena, CA 91125, USA
\item[\perugia] INFN-Sezione di Perugia and Universit\`a Degli 
     Studi di Perugia, I-06100 Perugia, Italy   
\item[\cmu] Carnegie Mellon University, Pittsburgh, PA 15213, USA
\item[\prince] Princeton University, Princeton, NJ 08544, USA
\item[\rome] INFN-Sezione di Roma and University of Rome, ``La Sapienza",
     I-00185 Rome, Italy
\item[\peters] Nuclear Physics Institute, St. Petersburg, Russia
\item[\potenza] INFN-Sezione di Napoli and University of Potenza, 
     I-85100 Potenza, Italy
\item[\riverside] University of Californa, Riverside, CA 92521, USA
\item[\salerno] University and INFN, Salerno, I-84100 Salerno, Italy
\item[\ucsd] University of California, San Diego, CA 92093, USA
\item[\sofia] Bulgarian Academy of Sciences, Central Lab.~of 
     Mechatronics and Instrumentation, BU-1113 Sofia, Bulgaria
\item[\korea]  Laboratory of High Energy Physics, 
     Kyungpook National University, 702-701 Taegu, Republic of Korea
\item[\alabama] University of Alabama, Tuscaloosa, AL 35486, USA
\item[\utrecht] Utrecht University and NIKHEF, NL-3584 CB Utrecht, 
     The Netherlands
\item[\purdue] Purdue University, West Lafayette, IN 47907, USA
\item[\psinst] Paul Scherrer Institut, PSI, CH-5232 Villigen, Switzerland
\item[\zeuthen] DESY, D-15738 Zeuthen, 
     FRG
\item[\eth] Eidgen\"ossische Technische Hochschule, ETH Z\"urich,
     CH-8093 Z\"urich, Switzerland
\item[\hamburg] University of Hamburg, D-22761 Hamburg, FRG
\item[\taiwan] National Central University, Chung-Li, Taiwan, China
\item[\tsinghua] Department of Physics, National Tsing Hua University,
      Taiwan, China
\item[\S]  Supported by the German Bundesministerium 
        f\"ur Bildung, Wissenschaft, Forschung und Technologie
\item[\ddag] Supported by the Hungarian OTKA fund under contract
numbers T019181, F023259 and T024011.
\item[\P] Also supported by the Hungarian OTKA fund under contract
  numbers T22238 and T026178.
\item[$\flat$] Supported also by the Comisi\'on Interministerial de Ciencia y 
        Tecnolog{\'\i}a.
\item[$\sharp$] Also supported by CONICET and Universidad Nacional de La Plata,
        CC 67, 1900 La Plata, Argentina.
\item[$\diamondsuit$] Also supported by Panjab University, Chandigarh-160014, 
        India.
\item[$\triangle$] Supported by the National Natural Science
  Foundation of China.
\end{list}
}
\vfill


\begin{table}
  \vspace{0.2cm}
  \begin{center}
    \begin{tabular}{|c||c||c|c|c|c|} \hline
 $W_\mathrm{vis}$ (GeV) & $N_\mathrm{data}$ &$N_\mathrm{c \bar{c}}$  &$N_\mathrm{uds}$ & $N_\mathrm{b}$ & $N_\mathrm{bkgd}$ \\ \hline
 $\phantom{0}3-5\phantom{0}$  & 261    & 208 & \phantom{0}48  & 0.2   & 4.6 \\ 
 $\phantom{0}5-10$ & 741   & 519 & 186 & \phantom{.}31  & 5.3 \\ 
 $10-15$           & 527  & 360 & 116 & \phantom{.}50  & 1.3 \\ 
 $15-25$           & 541  & 352 & 127 & \phantom{.}61  & 1.1 \\ 
 $25-40$           & 276  & 183 & \phantom{0}55  & \phantom{.}36  & 2.3 \\ 
 $40-70$           & \phantom{0}85  & \phantom{0}47  & \phantom{0}22  & \phantom{.}12  & 3.8 \\ 
 $\phantom{00}>70$ & \phantom{00}3     & \phantom{.}0.3   & \phantom{00}2   & 0.5   & 0.2 \\ \hline
    \end{tabular}
  \end{center}
  \caption{The number of events for the data, $N_\mathrm{data}$, the 
expected signal, $N_\mathrm{c \bar{c}}$, the background 
from light flavours, $N_\mathrm{uds}$, b production $N_\mathrm{b}$ 
and other processes $N_\mathrm{bkgd}$ as a function of $W_\mathrm{vis}$.}
  \label{tab:numbevents} 
\end{table}

\begin{table}
  \vspace{0.2cm}
  \begin{center}
    \begin{tabular}{|c||c|c|c|c|} \hline
 $\Delta W_\mathrm{\gamma\gamma}$ (GeV) & $N_\mathrm{c \bar{c}}$ unfolded  & Efficiency (\%) & $\mathrm{\Delta \sigma (e^+ e^- \rightarrow e^+ e^- c \bar{c} X})$ (nb) & $\mathrm{\sigma (\gamma\gamma \rightarrow c \bar{c} X})$ (nb) \\ \hline
 \phantom{0}$5-10$ & 428  & 0.33 $\pm$ 0.01  & 0.316   $\pm$ 0.013 $\pm$ 0.040  & 21.0  $\pm$ 0.8 $\pm$ 2.9   \\ 
 $10-15$ & 268  & 0.56 $\pm$ 0.02  & 0.117   $\pm$ 0.004 $\pm$ 0.015 & 18.6  $\pm$ 0.7 $\pm$ 2.6   \\ 
 $15-25$ & 389  & 0.70 $\pm$ 0.02  & 0.136   $\pm$ 0.005 $\pm$ 0.018 & 24.2  $\pm$ 0.9 $\pm$ 3.4   \\ 
 $25-40$ & 325  & 0.72 $\pm$ 0.03  & 0.110   $\pm$ 0.004 $\pm$ 0.015 & 33.0  $\pm$ 1.3 $\pm$ 4.9   \\ 
 $40-70$ & 196  & 0.57 $\pm$ 0.03  & 0.084   $\pm$ 0.004 $\pm$ 0.013 & 38.8  $\pm$ 1.9 $\pm$ 6.2  \\ \hline
    \end{tabular}
  \end{center}
  \caption{Unfolded number of charm events, $N_\mathrm{c \bar{c}}$, efficiencies
and cross section values as a function of $W_\mathrm{\gamma\gamma}$. 
The first uncertainty on the cross section is statistical and the second is systematic. 
The cross section values are evaluated at the centre of each $W_\mathrm{\gamma\gamma}$ interval.}
  \label{tab:cross} 
\end{table}

\begin{table}
  \vspace{0.2cm}
  \begin{center}
    \begin{tabular}{|l||c|c|c|c|c|c|} \hline
\multicolumn{1}{|c||}{} & \multicolumn{5}{c|}{$W_\mathrm{\gamma\gamma}$} \\ \hline
\multicolumn{1}{|c||}{Source of uncertainty} & $5-10$ GeV &$10-15$ GeV &$15-25$ GeV &$25-40$ GeV & $40-70$ GeV \\ \hline \hline
Charm purity & 11.3 & 12.2 & 12.2 & 12.6 & 12.6  \\ 
Unfolding & \phantom{0}5.0 & \phantom{0}5.0 & \phantom{0}5.0 & \phantom{0}5.0 & \phantom{0}5.0  \\ 
Charm efficiency & \phantom{0}3.8 & \phantom{0}3.7 & \phantom{0}3.3 & \phantom{0}3.9 & \phantom{0}5.0  \\ 
$\sigma(\mathrm{e^+ e^- \rightarrow e^+ e^- b \bar{b} X})$  & \phantom{0}3.5 & \phantom{0}1.6 & \phantom{0}2.8  & \phantom{0}3.4 & \phantom{0}4.1  \\ 
Photon flux  & \phantom{0}0.1 & \phantom{0}1.0 & \phantom{0}1.5 & \phantom{0}2.7 & \phantom{0}4.8  \\ 
Trigger efficiency & \phantom{0}2.1 & \phantom{0}2.1 & \phantom{0}2.2 & \phantom{0}2.3 & \phantom{0}2.3   \\ \hline
Total & 13.6 & 14.0  & 14.1  & 14.9 & 15.9  \\ \hline
    \end{tabular}
  \end{center}
  \caption{Systematic uncertainties (in percent)
on the cross section for the process 
$\mathrm{\gamma \gamma \rightarrow c \bar{c} X}$.}
  \label{tab:systematics} 
\end{table}

\begin{table}
  \vspace{0.2cm}
  \begin{center}
    \begin{tabular}{|c||c|c|c|c|c|} \hline
 $W_\mathrm{\gamma\gamma}$ (GeV) & $5-10$  & $10-15$ & $15-25$ & $25-40$ & $40-70$ \\ \hline
 \phantom{0}$5-10$ & 1.0\phantom{00}    &       &        &        &       \\ 
 $10-15$           & 0.739  & 1.0\phantom{00}   &        &        &       \\ 
 $15-25$           & 0.294  & 0.703 & 1.0\phantom{00}    &        &       \\ 
 $25-40$           & 0.090  & 0.302 & 0.757  & 1.0\phantom{00}    &       \\ 
 $40-70$           & 0.014  & 0.093 & 0.388  & 0.778  & 1.0\phantom{00}   \\ \hline
    \end{tabular}
  \end{center}
  \caption{Correlation matrix of the data after unfolding.}
  \label{tab:correlation} 
\end{table}


\newpage
\clearpage

\newpage

\newpage

\newpage

\end{document}